\documentclass{aastex631}
\usepackage{amsmath}

\shorttitle{Pickup ion distributions}
\shortauthors{Du et al.}
\graphicspath{{./}{figures/}}

\begin{document}

\title{On the anisotropic velocity distribution of newborn pickup ions in the heliosheath}

\correspondingauthor{Senbei Du}
\email{sdu@bu.edu}

\author[0000-0003-1134-3909]{Senbei Du}
\affiliation{Center for Space Physics, Boston University,
Boston, MA 02215, USA}

\author[0000-0002-8767-8273]{Merav Opher}
\affiliation{Center for Space Physics, Boston University,
Boston, MA 02215, USA}
\affiliation{Astronomy Department, Boston University,
Boston, MA 02215, USA}

\author[0000-0002-3479-1766]{Marc Kornbleuth}
\affiliation{Center for Space Physics, Boston University,
Boston, MA 02215, USA}

\begin{abstract}

The evolution of the velocity distribution of pickup ions is crucial for understanding the energetic neutral atom (ENA) fluxes observed by Interstellar Boundary Explorer (IBEX).
Pickup ions in the heliosheath contain two main components: those transmitted across the heliospheric termination shock and those locally created within the heliosheath.
In this work, we discuss the velocity distribution of the latter locally created component.
We find that pickup ions created by the charge exchange of neutral solar wind may be a significant source of the observed ENA fluxes between about 100 eV and 1 keV.
Moreover, newborn pickup ions can maintain highly anisotropic velocity distribution in the heliosheath.
This is because the kinetic instabilities are weak after the solar wind flow decelerates at the termination shock.
Hybrid kinetic simulations show the mirror instability to be the dominant mode for conditions in the heliosheath close to the termination shock.
We estimate that effects of neutral solar wind and anisotropy may enhance the expected phase space density of newborn pickup ions by more than a factor of 100.

\end{abstract}


\section{Introduction} \label{sec:intro}

Pickup ions are created throughout the heliosphere from ionized neutral atoms.
Newborn pickup ions typically follow a ring-beam distribution, which is known to drive various kinetic instabilities such as ion cyclotron, mirror, ion Bernstein, etc. \citep[e.g.,][]{Wu1972,Lee1987,Summerlin2014,Florinski2016,Min2018}.
The waves associated with the instabilities scatter the pickup ions in pitch angle.
In the supersonic solar wind, it is generally assumed that interstellar pickup ions (whose parent neutral atoms are of interstellar origin) generate Alfv\'en ion cyclotron waves, which rapidly scatter the pickup ions into the bispherical shell distribution \citep[e.g.,][]{Lee1987,Williams1994,Isenberg1996}, and the scattering happens on a timescale significantly shorter than the ionization timescale.
The bispherical shell is approximately isotropic when $u_{sw} \gg V_A$, where $u_{sw}$ is the solar wind speed and $V_A$ is the Alfv\'en speed.
Thus, it is justified to use an isotropic model to describe the interstellar pickup ions in the supersonic solar wind \citep[e.g.,][]{Vasyliunas1976,Zhao2019,McComas2021}.
Pickup ions are also important in planetary or cometary exospheres \citep[e.g.,][]{Gary1988,Cheng2023,Cheng2024} and the local interstellar medium \citep[e.g.,][]{Liu2012,Mousavi2020}.
In all these situations, the rapid instability growth and isotropization of ion distributions are found.

Comparatively, the pickup ions in the heliosheath are much less understood.
The heliosheath is the region where the solar wind decelerates after crossing the heliospheric termination shock.
It is important to distinguish two pickup ion components that are present in the heliosheath as they have very different characteristics.
The first component is the pickup ions created in the supersonic solar wind and transmitted across the termination shock (denoted as transmitted pickup ions).
The second component is the pickup ions locally created in the heliosheath, mainly by the charge exchange between ions and neutrals.
We focus on the second pickup ion component in this work.
The slower flow velocity downstream of the termination shock means that less free energy is contained in the heliosheath-created pickup ions, which leads to weaker instabilities and pitch angle scattering.
Consequently, an anisotropic velocity distribution may be expected for the heliosheath pickup ions.
Furthermore, the plasma just downstream of the termination shock has a high beta (ratio between thermal and magnetic pressure) $\beta \gtrsim 20$ as the heated pickup ions transmitted through the termination shock become the energetically dominant component.
The high-beta environment is known to drive the mirror instability preferentially, in contrast to the ion cyclotron instability in the supersonic solar wind \citep{Yoon2017}.

Understanding the pickup ions in the heliosheath is crucial for modeling the energetic neutral atom (ENA) fluxes measured by IBEX (Interstellar Boundary Explorer) and Cassini/INCA (Ion and Neutral Camera).
As the pickup ions charge exchange with neutrals, they become a source of the observed ENAs.
It has been suggested that the modeled ENA fluxes underpredict the observations by IBEX-Lo by more than an order of magnitude at energies $\sim 100$ eV in all directions \citep{Galli2023}.
Energy diffusion due to turbulence may partially explain the discrepancy, especially in the tail regions \citep{Fahr2016,Zirnstein2018,Zirnstein2018b}, but our recent work suggests that heliosheath turbulence is likely not strong enough to be main cause of the ENA gap in Voyager directions \citep{Du2024}. A recent ENA model also shows that even strong energy diffusion cannot fully explain the ENA gap in the upwind direction \citep{Baliukin2024}.
However, ENA models assume all the ions to be isotropic, which may not be true especially for the heliosheath pickup ions.
We note that some models of the IBEX ``ribbon'' \citep{McComas2009}, i.e., weak scattering, do rely on anisotropic pickup ions in the interstellar medium, but they do not concern the heliosheath ions.
The anisotropy of the heliosheath ions may contribute to the ENA gap if the anisotropy enhances the part of the phase-space density that gives birth to the ENAs {detectable at 1 au}.
Our analysis suggests that this is indeed the case.
In addition to the anisotropy, the secondary ENA mechanism is also commonly ignored for the models of low-energy ENAs that constitute the globally distributed flux.
The secondary ENA mechanism is believed to be the key to explaining the IBEX ribbon, which is seen most prominently in high-energy ENAs ($\sim$ 1--4 keV) by IBEX-Hi.

In this paper, we investigate the anisotropic velocity distribution of newborn pickup ions in the heliosheath immediately downstream of the termination shock.
The results have important implications for modeling observed ENA fluxes.
The effects of secondary ENAs are included in this work, as described in Section \ref{sec:secondary}.
Hybrid kinetic simulation results are presented in Section \ref{sec:hybrid}.
Conclusions and discussions are found in Section \ref{sec:conclusion}.

\section{Secondary ENAs from the heliosheath}\label{sec:secondary}

The charge exchange between the supersonic solar wind protons and interstellar neutral hydrogen atoms generates neutral solar wind atoms that travel radially outward from the Sun.
Downstream of the heliospheric termination shock, charge exchange can happen between the neutral solar wind and the shocked solar wind ions (including transmitted thermal and pickup ions), which produces new pickup ions in the heliosheath.
ENAs that can be observed by IBEX near Earth are produced by a third charge exchange between the new pickup ions and the neutrals in the heliosheath.
This triple charge-exchange process is known as the secodary ENA mechanism.
If the second and third charge exchanges occur in the local interstellar medium instead of the heliosheath, the secondary ENA mechanism is widely thought to give rise to the IBEX ribbon due to the draped interstellar magnetic field outside the heliopause.

The effects of secondary ENAs are generally ignored in the heliosheath because the neutral solar wind density is very low compared to the interstellar neutrals (including the hotter hydrogen wall neutrals).
For example, \citet{Heerikhuisen2016} showed that the interstellar neutral density ($\sim 0.1 \mathrm{cm}^{-3}$) is about 500 times higher than the neutral solar wind density ($\sim 0.0002 \mathrm{cm}^{-3}$) just downstream of the termination shock.
Thus, the new pickup ions created in the heliosheath are much more likely to originate from the interstellar neutrals \citep[e.g.,][]{Baliukin2020}.
However, the interstellar neutrals are nearly stationary in the solar inertial frame, with a bulk velocity of only $\sim 20$ km/s, meaning that the newborn pickup ions due to the interstellar neutrals are not very likely to be observed near Earth by IBEX without energy diffusion.
On the other hand, the newborn pickup ions due to the neutral solar wind can have a higher radially inward velocity, and their relative contribution to the IBEX ENA maps will be boosted.

\begin{figure}[ht!]
\centering
\includegraphics[width=0.5\linewidth]{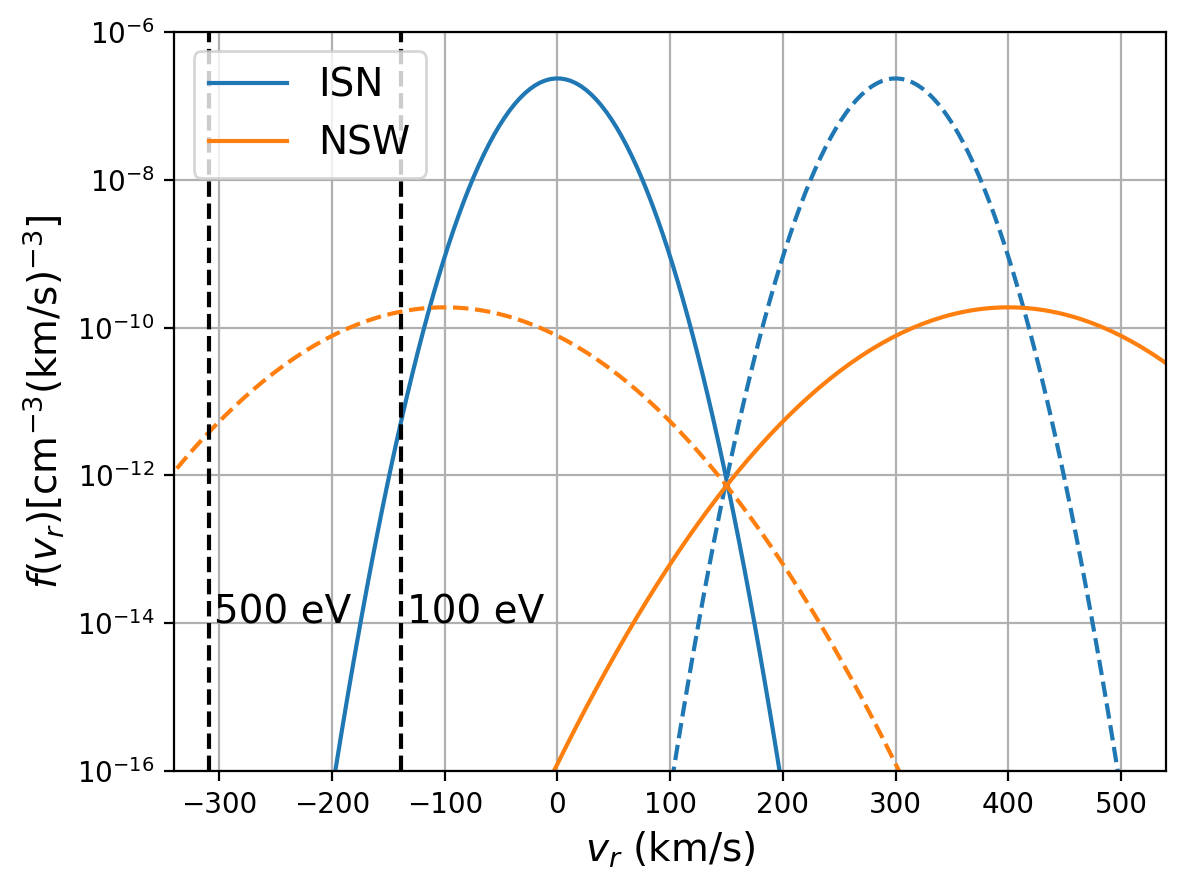}
\caption{The radial velocity distribution function of interstellar neutrals and the neutral solar wind. The distributions are mirrored with respect to the bulk flow velocity of 150 km/s, plotted as the dashed lines of the corresponding colors. \label{fig:neutral}}
\end{figure}

A more quantitative estimate can be obtained by comparing the radial velocity distribution of the interstellar neutrals and the neutral solar wind.
We assume both neutral components are bi-Maxwellian distributed, i.e.,
\[ f_M(\boldsymbol{v}) = \frac{n}{(2\pi)^{3/2} v_{th} v_{th\perp}^2} \exp\left[-\frac{(v_r-v_0)^2}{2 v_{th}^2} - \frac{|\boldsymbol{v}_{\perp}|^2}{2 v_{th\perp}^2}\right], \]
where $n$ is the number density; $v_{th}$ is the radial thermal velocity, $v_{th\perp}$ is the perpendicular thermal velocity, and the velocity vector $\boldsymbol{v}$ contains the radial component $v_r$ and the component perpendicular to radial $\boldsymbol{v}_{\perp}$.
The radial flow velocity is $v_0$ and the perpendicular bulk flow is neglected.
Figure \ref{fig:neutral} plots the distribution function for $\boldsymbol{v}_{\perp} = \boldsymbol{0}$.
The interstellar neutrals (ISN) have a number density $n = 0.1 \mathrm{cm}^{-3}$, flow velocity $v_0 = 0$, and thermal velocity $v_{th} = v_{th\perp}$ 30 km/s, plotted as the blue solid line.
The neutral solar wind (NSW) has a number density $n = 0.0002 \mathrm{cm}^{-3}$, flow velocity $v_0 = 400$ km/s, and thermal velocity $v_{th} = 80$ km/s, $v_{th\perp} = 30$ km/s, plotted as the orange solid line.
Since we are interested in the pickup ions instead of the neutrals themselves, the effective thermal velocity $v_{th}$ here includes the combination of the actual thermal velocity of the neutrals $\delta v_{th}$ and the turbulent fluctuations of the radial solar wind flow velocity in the heliosheath $\delta v_{turb}$.
The actual thermal velocity is estimated based on the results by \citet{Heerikhuisen2016}: $\delta v_{th} \sim 15$ km/s for the interstellar neutrals (dominated by the hydrogen wall population) and $\delta v_{th} \sim 75$ km/s for the neutral solar wind.
The high temperature of the neutral solar wind is mainly due to the variation in the solar wind flow velocity throughout the heliosphere.
The turbulent velocity fluctuations are estimated from Voyager observations \citep{Richardson2008,Du2024}: $\delta v_{turb} \sim 26$ km/s.
The effective thermal velocity used in Figure \ref{fig:neutral} is then estimated by $v_{th} = \sqrt{\delta v_{th}^2 + \delta v_{turb}^2} \simeq 30$ km/s (ISN) and $\simeq 80$ km/s (NSW).
We note that the NSW is assumed to be anisotropic with $v_{th\perp} > v_{th}$, unlike the ISN.
This is expected because the variation in solar wind velocity is predominantly in the radial direction.
For example, near 1 au, the radial solar wind commonly varies between 300 and 800 km/s while the nonradial velocity is usually below 50 km/s.

After a pickup ion is created, it gyrates about the magnetic field and is advected by the solar wind flow.
Therefore, to represent the relative contribution of the two neutral populations to pickup ions, we mirror the neutral distribution function about the heliosheath flow velocity, assumed to be 150 km/s.
This is represented by the blue and orange dashed lines in Figure \ref{fig:neutral} (for interstellar neutrals and the neutral solar wind).
The peaks of the NSW pickup ions are located at $v_r = -100$ km/s and 400 km/s, and the peaks of the ISN pickup ions are located at $v_r = 0$ and $300$ km/s.
As a reference, the velocities that correspond to 100 and 500 eV are plotted as the vertical dashed lines.
We note that only negative velocities are relevant for ENA observations because ENAs created in the outer heliosphere need to propagate radially inward to reach the detectors near Earth \citep{Fuselier2021}.
Figure \ref{fig:neutral} demonstrates that at energies about 100 eV or higher, the NSW pickup ions become the dominant contributor despite the very low density of the neutral solar wind.
It is difficult to quantify the exact enhancement factor of the modeled ENA fluxes produced by NSW pickup ions because the results are somewhat sensitive to the assumptions about the neutral distribution and its spatial variation, as well as the transmitted ions.
We nevertheless provide an estimate of $f_{NSW+ISN}/f_{ISN} \sim$ 30 (which is the enhancement in the neutral source distribution for pickup ions created in the heliosheath) around 100 eV based on Figure \ref{fig:neutral}.
More precise calculation of the neutral distribution requires kinetic neutral modeling, which will be deferred to a future work.

\section{Hybrid kinetic simulation} \label{sec:hybrid}
\subsection{Simulation setup}

In this section, we demonstrate how the ion distribution evolves in the presence of newborn pickup ions in the heliosheath via hybrid kinetic simulations (or just hybrid simulations), which consider ions as macroparticles and treat electrons as a charge-neutralizing fluid.
Hybrid simulations are suitable for our study because we are mostly interested in the ion kinetic physics.
We use the open-source code Hybrid-VPIC \citep{Le2021,Bowers2008}.

We consider a 2D homogeneous and periodic box of the size $L_x = L_z = 256 d_i$, with a grid resolution of $512^2$.
The background magnetic field is in the $x$-direction.
We use $10^4$ macroparticles per cell per ion species, which are sufficient to ensure that the numerical noise does not overwhelm the physical instability.
The simulations include three ion species: a Maxwellian component that represents the thermal solar wind ions; a hotter Maxwellian component that represents pickup ions transmitted through the termination shock; and a ring component that represents the newborn heliosheath pickup ions.
We focus on the evolution of the newborn pickup ions in this study, and we do not expect the results to depend critically on the distribution of transmitted ions, though they may play a minor role in the quantitative calculation of the linear growth rates in Section \ref{sec:time}.
The number density ratio is $n_t / n_p = 3/1$ between thermal and transmitted pickup ions.
The temperature is $kT = 0.75 m_p V_A^2$ for the thermal ions and 30 $m_p V_A^2$ for the transmitted pickup ions, where $V_A$ is the Alfv\'en speed, $k$ is the Boltzmann constant, and $m_p$ is the proton mass.
The velocity in the simulations are translated to real units assuming $V_A = 50$ km/s.
The parameters are consistent with the expected ion distribution downstream of the heliospheric termination shock in the Voyager 2 direction, as suggested by previous simulations \citep{Giacalone2021}.

\begin{figure}[!ht]
\centering
\includegraphics[width=0.5\linewidth]{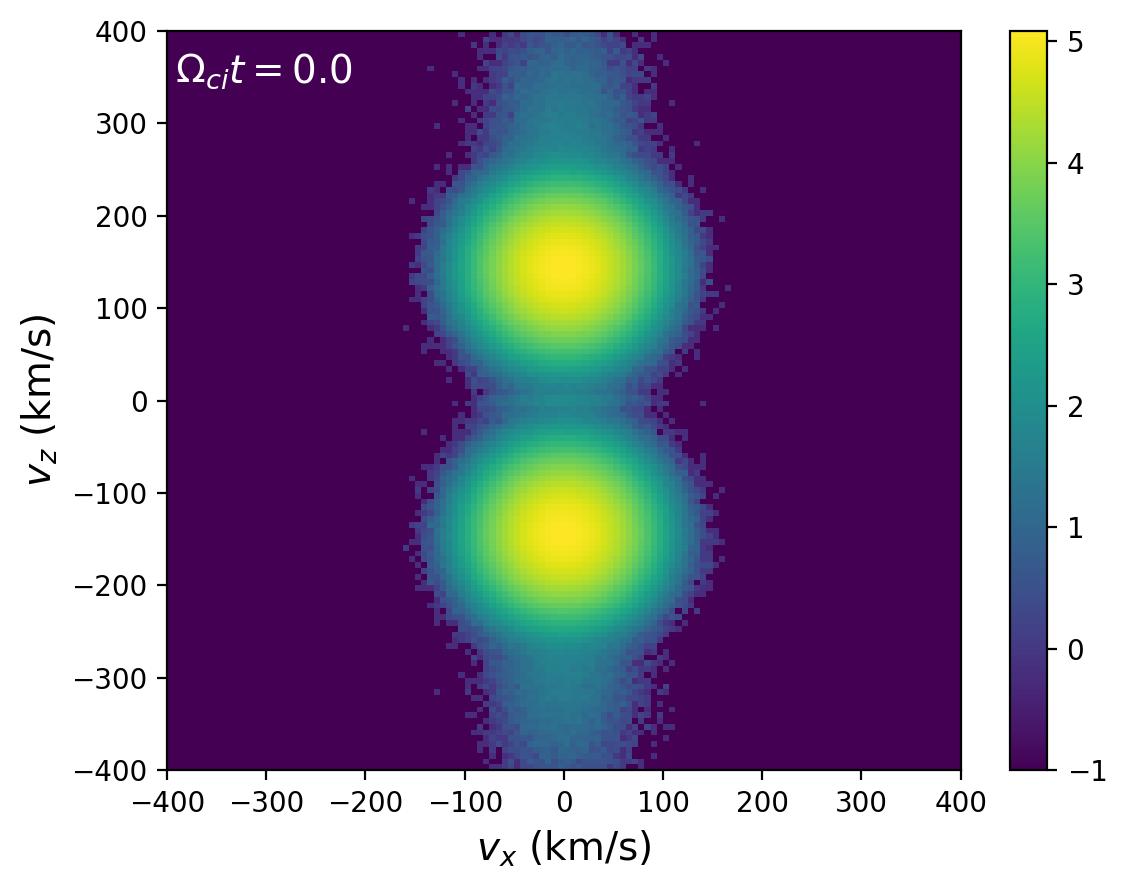}%
\includegraphics[width=0.5\linewidth]{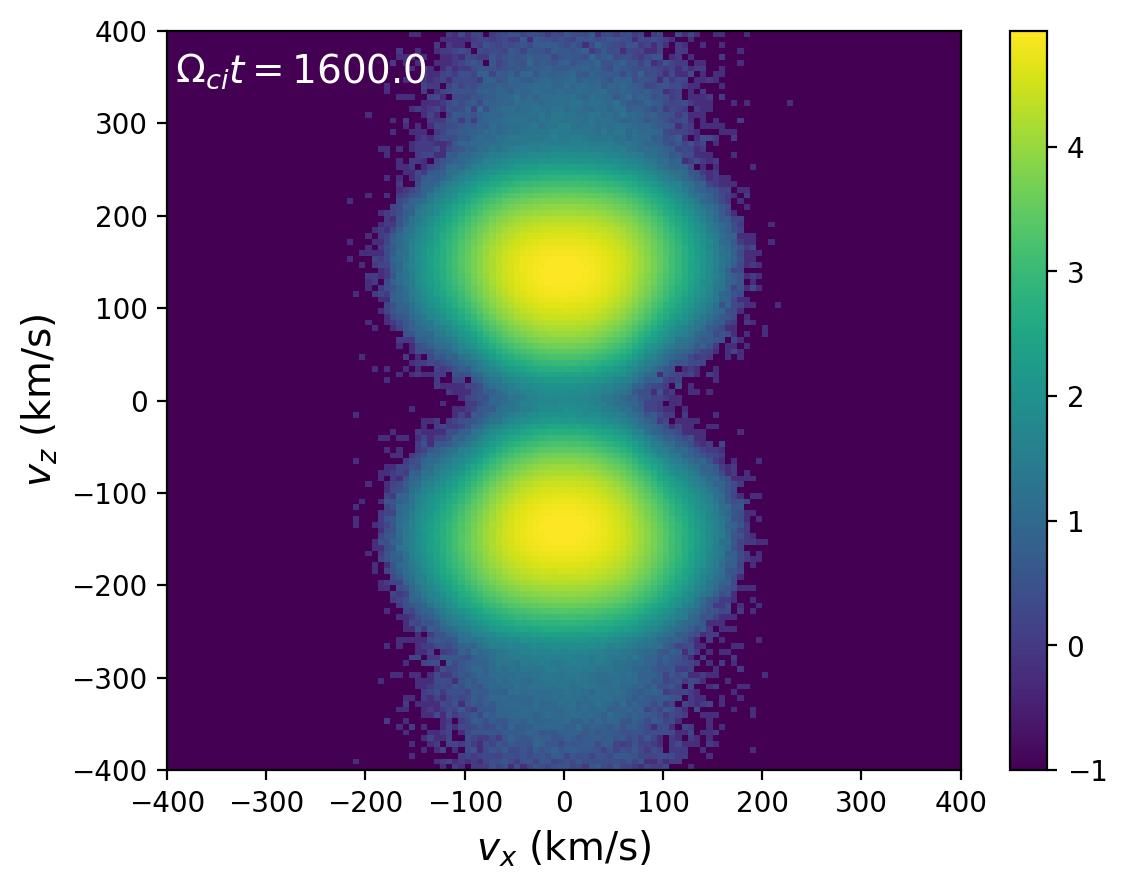}
\caption{The common logarithm of initial velocity distribution (unnormalized) of newborn pickups at $v_y \simeq 0$. The left panel is the initial distribution and the right panel is taken at time $\Omega_{ci}t \simeq 1600$.}\label{fig:init}
\end{figure}

We assume the newborn heliosheath pickup ions follow a ring distribution initially. The velocity distribution function is expressed as
\begin{equation}\label{eq:fr}
  f(v_{\parallel}, v_{\perp}) = \frac{A}{v_\perp} \exp\left[-\left(\frac{v_{\parallel}}{\delta v_{\parallel}}\right)^2\right] \exp\left[-\left(\frac{v_{\perp} - v_r}{\delta v_{\perp}}\right)^2\right].
\end{equation}
The ring velocity $v_r$ is determined by the flow velocity perpendicular to the magnetic field.
We set $v_r = 3 V_A$ in our simulations based on the flow velocity of $\sim$ 150 km/s measured by Voyager 2 downstream of the termination shock.
Additionally, the pickup ions created from the neutral solar wind are also included with $v_r = 5 V_A$ according to the relative velocity between NSW and the heliosheath flow, and they account for 0.2\% of the newborn pickup ion density.
The distribution is broadened as Gaussian functions, parameterized by the widths $\delta v_{\parallel}$ and $\delta v_{\perp}$.
We set $\delta v_{\perp} = 0.6 V_A$ for the ISN pickup ions, $\delta v_{\perp} = 1.5 V_A$ for the NSW pickup ions, and $\delta v_{\parallel} = 0.6 V_A$ for both.
These parameters are consistent with our estimated neutral distribution from the previous section.
The initial velocity distribution (unnormalized) of newborn pickup ions at $v_y \simeq 0$ is plotted in the left panel of Figure \ref{fig:init}.
The constant $A$ is a normalization factor:
\[ \int_{0}^{\infty} \int_{-\infty}^{\infty} f(v_{\parallel}, v_{\perp}) 2\pi v_{\perp}dv_{\perp} dv_{\parallel} = n_n \quad\Rightarrow\quad A = \frac{n_n}{\pi^{2} \delta v_{\perp} \delta v_{\parallel} \mathrm{erfc} (-v_r / \delta v_\perp)}, \]
where $n_n$ is the newborn pickup ion number density and erfc($\ldots$) is the complementary error function.
We note that the distribution function is slightly different from the form considered by several previous work \citep[e.g.,][]{Florinski2016,Min2018} as we include $v_{\perp}$ in the denominator so that the reduced distribution $f_r \equiv 2\pi v_{\perp} f$ is symmetric in $v_{\perp}$ with respect to $v_r$.
This is more reasonable than the other choice of omitting $v_{\perp}$ in the denominator---which makes the distribution itself symmetric---because filling up the gyrophase should preserve the symmetry in the reduced distribution, assuming the parent neutrals (including both ISN and NSW) follow a symmetric Maxwellian distribution.
Two simulation runs are performed, with newborn pickup ion density ratios of $n_n/n_i\simeq 7$\% and 3.5\%, where $n_i = n_t + n_p + n_n$ is the total ion density.
Unless otherwise indicated, most of the results are shown for the run with 7\% pickup ion.
The simulation is run for $\Omega_{ci} t \sim 1600$ (and 2400 for the 3.5\% run), where the cyclotron timescale $\Omega_{ci}^{-1} = m_p c/eB$ is about 100 s in the heliosheath (for $B \simeq 0.1$ nT).

\subsection{Simulation results}

\begin{figure}[!ht]
\centering
\includegraphics[width=0.5\linewidth]{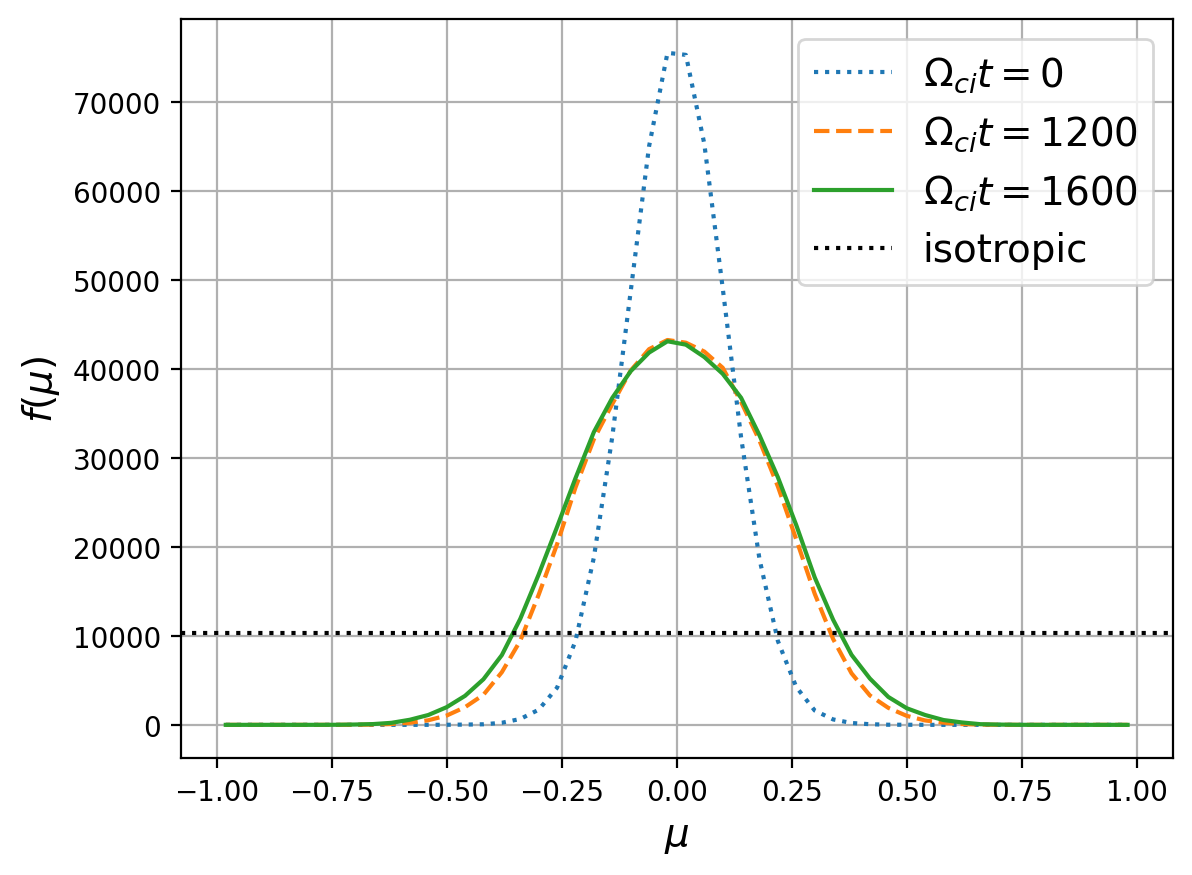}%
\includegraphics[width=0.5\linewidth]{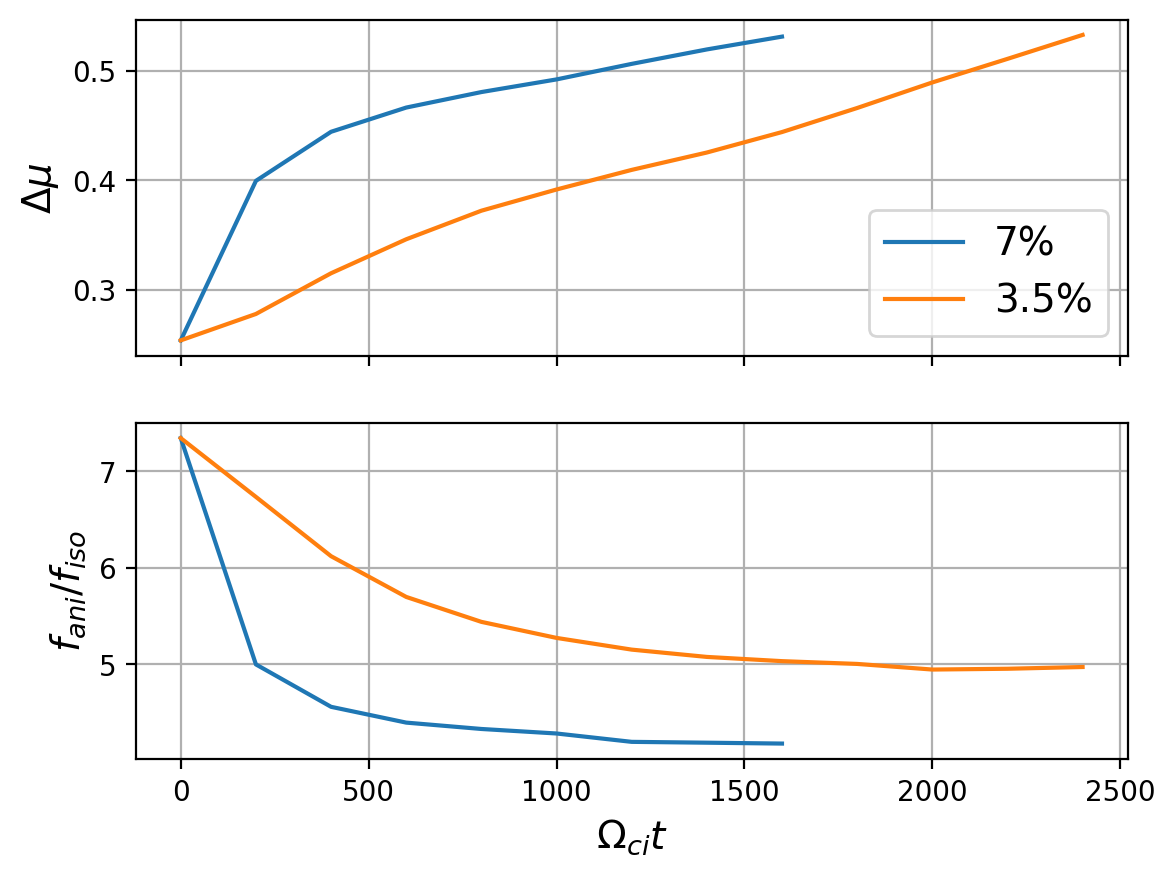}
\caption{Left: unnormalized pitch angle (cosine) distribution for selected particles at several times of the simulation. Top right: evolution of the width of the pitch angle distribution. Bottom right: enhancement of the phase-space density near $\mu = 0$ over an isotropic distribution.} \label{fig:PA}
\end{figure}

The main result of our simulation is that the newborn pickup ions are not scattered to isotropy.
The anisotropic velocity distribution at the end of the simulation is illustrated in the right panel of Figure \ref{fig:init}.
Motivated by IBEX observations, we focus on 100 eV ENAs, which come from pickup ions with velocity $v_{\perp}$ around 290 km/s in the simulation frame.
To illustrate the evolution of anisotropy, particles with velocities 280 km/s $\le |v| = \sqrt{v_x^2 + v_y^2 + v_z^2} \le$ 300 km/s are selected.
A direct comparison with IBEX observations needs to consider the specific energy bands and instrument response functions \citep{Fuselier2009}, which is beyond the scope of this work.
The cosine pitch angle of a particle is computed as $\mu = \cos\theta = v_x / |v|$.
The left panel of Figure \ref{fig:PA} shows the unnormalized pitch angle distribution $f(\mu)$ from the simulation.
The initial distribution is plotted as the blue dotted line, and the distributions at two other times ($\Omega_{ci}t = $1200 and 1600) are plotted as the orange dashed line and green solid line.
As a reference, the isotropic distribution corresponds to a horizontal line, plotted as the black dotted line.
The pitch angle distribution peaks around 0, as expected from a ring distritution of the newborn pickup ions.
The pickup ions are then scattered by self-generated waves, resulting in the broadening of the pitch angle distribution later in the simulation.
However, complete isotropization does not occur at the end of the simulation.
We quantify the broadening by the full width of the distribution at half-maximum, $\Delta \mu$.
The top-right panel of Figure \ref{fig:PA} shows the evolution of the width.
Since observed ENAs are expected to come from ions with $\theta \simeq 90^{\circ}$, the effect of the anisotropy on modeled ENA flux can be estimated by the enhancement factor of the phase-space density near $\mu = 0$ relative to an hypothetical isotropic distribution initially.
This enhancement factor $f_{ani}/f_{iso}$ is plotted in the bottom-right panel of Figure \ref{fig:PA}.
It shows that the factor decreases from its initial value of $f_{ani}/f_{iso} \simeq 7$ and stablizes at $\sim 4.2$, although $\Delta \mu$ still appears to increase at the end.
This may be due to additional heating by wave--particle interaction, increasing the number of particles in the selected energy band (numerical heating is well under 0.1\% for the simulation).
We note that the enhancement factor due to anisotropy should be almost independent of the effects of neutral solar wind because the number density of NSW pickup ions is so low that they behave like test particles and do not affect the evolution of the system qualitatively.
The total enhancement in the ion phase-space density at $\mu \simeq 0$ over a model that does not include the effects of NSW and anisotropy can then be estimated by multiplying $f_{ani}/f_{iso}$ from Figure \ref{fig:PA} and $f_{NSW+ISN} / f_{ISN}$ from Figure \ref{fig:neutral}, yielding a factor of $\sim 120$ at $\sim 100$ eV.
Results from the 3.5\% run and 7\% run are qualitatively similar, and both are included in the right panels of Figure \ref{fig:PA}.
Since the instability is weaker with lower pickup ion density, the enhancement factor saturates at a higher level, $f_{ani}/f_{iso} \sim 5$, which should correspond to a slightly higher {total enhancement}.

\begin{figure}[!ht]
\centering
\includegraphics[width=0.32\linewidth]{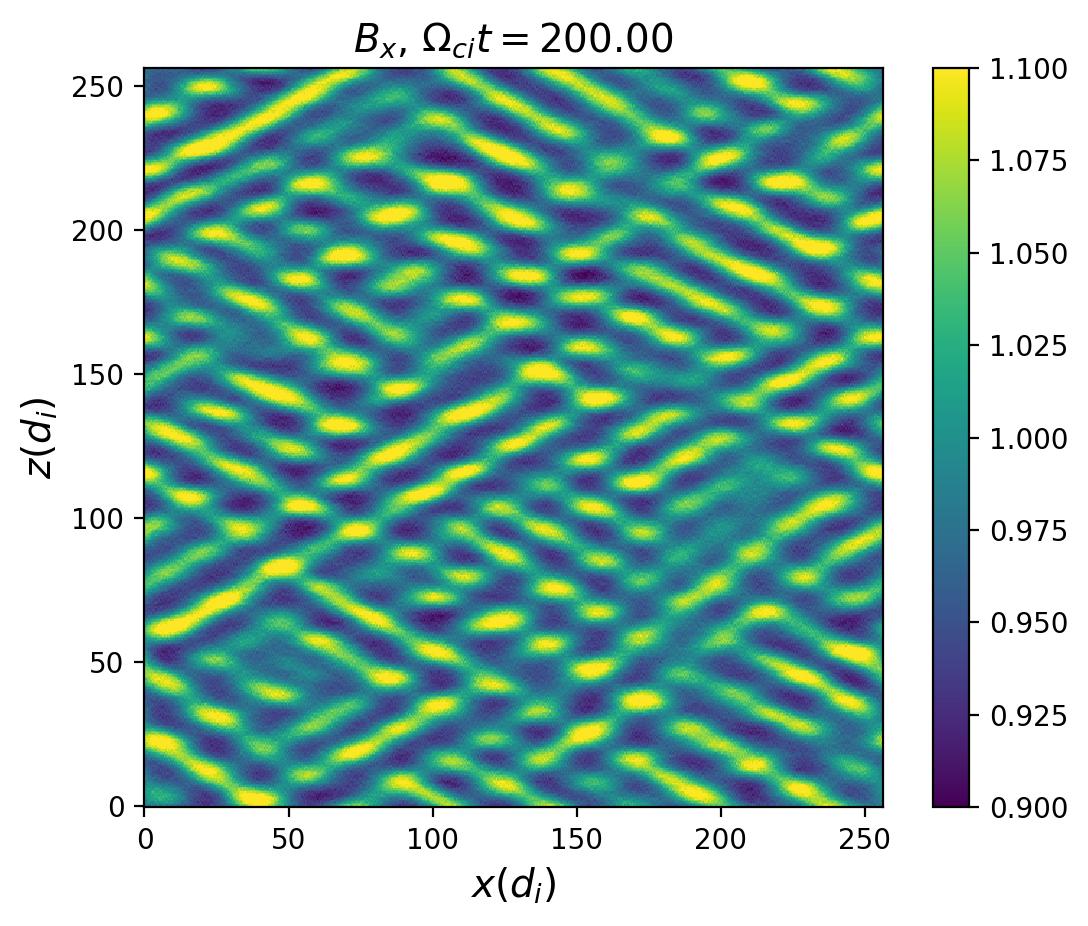}%
\includegraphics[width=0.32\linewidth]{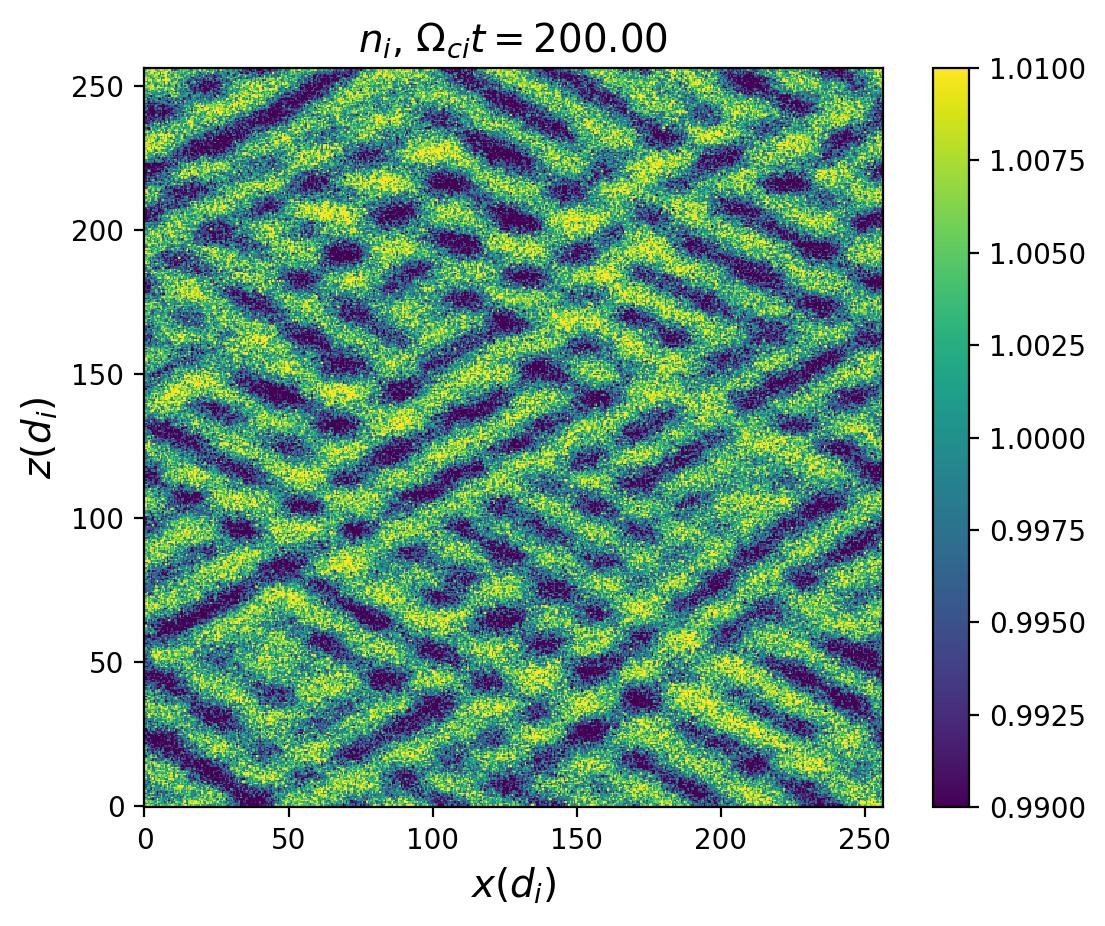}%
\includegraphics[width=0.36\linewidth]{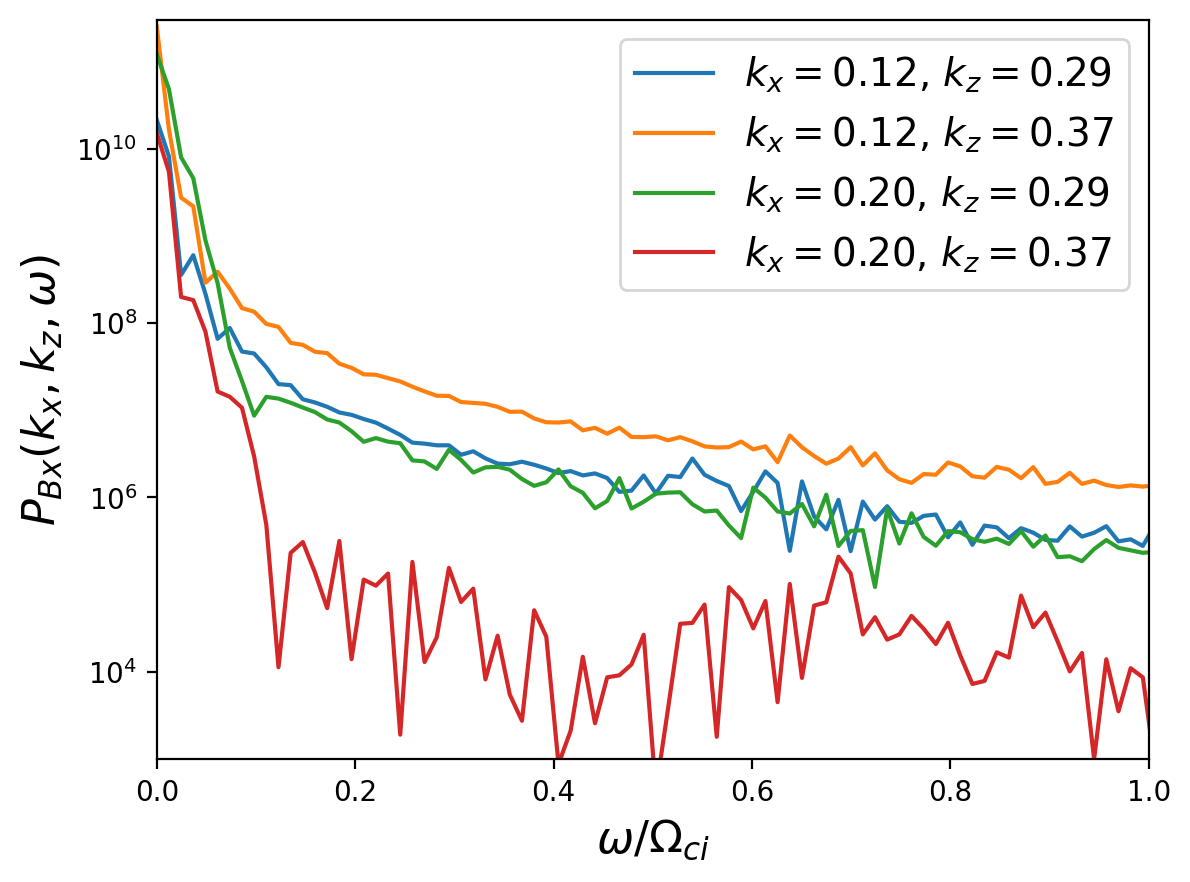}
\caption{Snapshots of magnetic field component $B_x$ (left panel) and ion density $n_i$ (middle panel) at $\Omega_{ci}t=200$ from the simulation. Right: frequency spectra of $\delta B_x$ at several selected wavevectors.}\label{fig:mirror}
\end{figure}

The scattering of newborn pickup ions is due to kinetic instabilities driven by the initial ring distribution.
We find that the mirror instability is dominant in our simulation, as demonstrated in Figure \ref{fig:mirror}.
The left and middle panels show the oblique structures in the compressive magnetic field component $B_x$ and total ion density $n_i$.
Another notable feature from the figure is the anticorrelation between magnetic field and density fluctuations (dark features in $B_x$ correspond to bright features in $n_i$), which is characteristic of mirror modes and slow magnetosonic waves.
To further understand the nature of the instability and fluctuations, we perform Fourier spectral analysis in both space and time for the fluctuation $\delta B_x$, obtaining the frequency--wavenumber spectral power $P_{Bx}(k_x, k_z, \omega)$.
The temporal Fourier transform is performed for simulation time between $\Omega_{ci}t = 100$ and 612.
Examples of the frequency spectra are shown in the right panel of Figure \ref{fig:mirror}.
We select four pairs of values for $k_x$ and $k_z$ where the fluctuations are strong.
The peak of each spectrum is located at $\omega = 0$, which is consistent with nonpropagating mirror mode structures.
The pitch angle scattering due to mirror instability tends to flatten the distribution in $v_{\parallel}$ \citep[e.g.,][]{Califano2008,Hellinger2017}, and this feature is exhibited in the right panel of Figure \ref{fig:init}.

\subsection{Comments on the instability timescale} \label{sec:time}

In principle, the newborn pickup ion number density should be determined by the charge-exchange process,
\begin{equation}\label{eq:np}
  \frac{n_{n}}{n_i} \simeq n_{H} \Delta v \sigma(\Delta v) \tau,
\end{equation}
where $n_H$ is the parent neutral hydrogen density, $\Delta v$ is the relative velocity between the parent neutral and the solar wind flow, $\sigma(\Delta v)$ is the charge-exchange cross section that depends on the relative velocity, and $\tau$ is the charge-exchange timescale.
Assuming $n_H = 0.1 \mathrm{cm}^{-3}$, $\Delta v = 150$ km/s, $\sigma(\Delta v) = 2\times 10^{-15} \mathrm{cm}^{2}$ \citep{Lindsay2005}, and $B = 0.1$ nT, we find that $\Omega_{ci}\tau \simeq 3\times 10^6(n_n/n_i)$.
Therefore, a pickup ion fraction of $n_n/n_i \simeq 0.07$ from our simulation corresponds to a charge-exchange timescale of $\Omega_{ci}\tau \simeq 2\times 10^5$, much longer than our simulation time of 1600.
A useful reference is the timescale for the solar wind to travel across the heliosheath $\tau_{sw} \simeq L/u_{sw} \sim 3\times 10^7$ s (or $\Omega_{ci}\tau_{sw} \sim 3\times 10^5$), considering the heliosheath thickness $L \sim 30$ au in the upwind direction \citep[e.g.,][]{Richardson2022}.
In this sense, our simulations have implicitly assumed an artificially fast charge-exchange timescale that matches the simulation run time, which is to keep the computational cost reasonable.
This is justifiable as an approximation because the change in plasma conditions is relatively insignificant over $\sim 10$--20 au, though the variation will become more notable over longer spatial scales (hundreds of astronomical units), which is especially important for the heliotail region.

On the other hand, the instability growth rate becomes smaller as the pickup ion fraction decreases.
Indeed, there is a minimum threshold for $n_n/n_i$ below which the instability will not grow at all.
Physically, this is because the instability due to newborn pickup ions must overcome the damping due to the thermal or transmitted pickup ions.
The linear growth rate of the mirror instability can be derived based on the particle distribution function.
Here, we consider the approximate dispersion relation by keeping the dominant terms in the susceptibility tensor \citep[e.g.,][]{Yoon1992,Yoon2017}:
\begin{eqnarray}
  \frac{\omega^2}{\omega_{pi}^2} &=& \frac{c^2k^2}{\omega_{pi}^2} - \sum_m \frac{n_m}{n_0} \zeta_{m0} Z(\zeta_{m0}) \lambda_m e^{-\lambda_m} [2I_0(\lambda_m) - 2I_1(\lambda_m)] \nonumber\\
  &+& \sum_r \frac{n_r}{n_0} \frac{2}{\sqrt{\pi}\delta v_{\perp}^3 \mathrm{erfc}(-v_r/\delta v_{\perp})} \int_0^{\infty} dv_\perp \left[2v_{\perp}(v_\perp - v_r) + \delta v_{\perp}^2\right] \left| J_1\left( \frac{k_\perp v_\perp}{\Omega_{ci}} \right) \right|^2 \exp \left[ -\frac{(v_\perp - v_r)^2}{\delta v_\perp^2} \right] \nonumber\\
  &-& \sum_r \frac{n_r}{n_0} \frac{4}{\sqrt{\pi}\delta v_{\perp} \delta v_{\parallel}^2 \mathrm{erfc}(-v_r/\delta v_{\perp})} \int_0^{\infty} dv_\perp v_{\perp}^2 \left| J_1\left( \frac{k_\perp v_\perp}{\Omega_{ci}} \right) \right|^2 \exp \left[ -\frac{(v_\perp - v_r)^2}{\delta v_\perp^2} \right] \left[ 1 + \frac{\omega}{k_\parallel \delta v_{\parallel}} Z\left( \frac{\omega}{k_\parallel \delta v_{\parallel}} \right) \right]. \label{eq:mirror}
\end{eqnarray}
The electron response is ignored since mirror mode is nearly nonpropagating and is only Landau resonant with ions.
The ions may include multiple isotropic Maxwellian components and ring components described by Equation \eqref{eq:fr}.
We note the slightly different expression from \citet{Min2018} and \citet{Mousavi2020} due to the different ring distribution that we use.
The subscript $m$ represents parameters for the Maxwellian components:
\[ \zeta_{m0} = \frac{\omega}{k_\parallel v_{tm}};\quad \lambda_m = \frac{k_\perp^2 v_{tm}^2}{2\Omega_{ci}^2}, \]
where $n_m$ and $v_{tm}$ are the number density and thermal velocity of each Maxwellian component; $\Omega_{ci}$ is the proton gyrofrequency.
Bessel function of the first kind $J_{1}$, modified Bessel function $I_{0,1}$, and the plasma dispersion function $Z$ are used in the dispersion relation.
The number density of each ring component is $n_r$.
The integration in perpendicular velocity is retained for the ring components, and it is to be computed numerically.

\begin{figure}[ht!]
\centering
\includegraphics[width=0.5\linewidth]{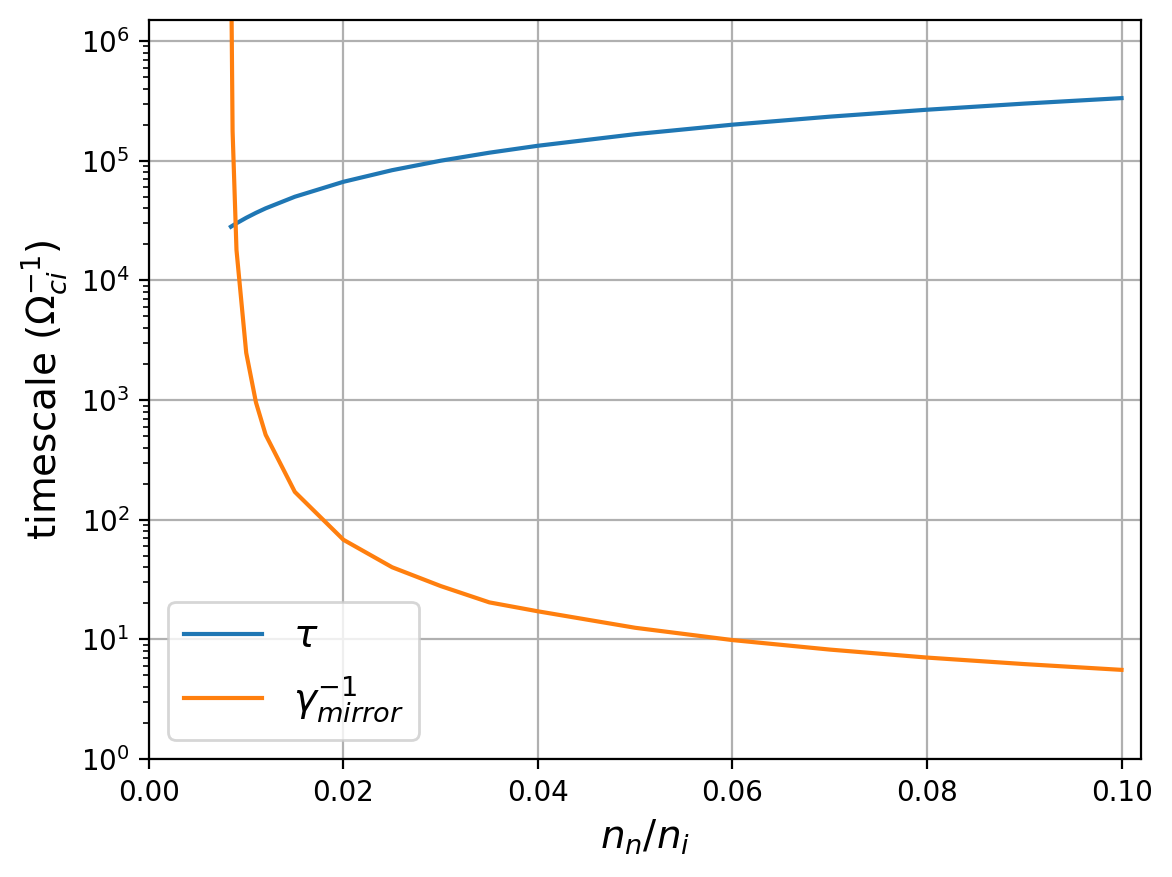}
\caption{The inverse maximum growth rates for the mirror mode $\gamma_{mirror}^{-1}$ (orange) and the charge-exchange timescale $\tau$ as functions of the newborn pickup ion fraction $n_n/n_i$. Both quantities are normalized by $\Omega_{ci}^{-1}$.\label{fig:time}}
\end{figure}

Equation \eqref{eq:mirror} is solved numerically, and the imaginary part of the frequency solution gives the instability growth rate $\gamma$. (The real frequency is approximately zero.)
The maximum growth rate is then found by searching in a grid of $k_{\perp}$ and $k_{\parallel}$.
Parameters that correspond to our simulation are used, including two Maxwellian components (thermal and transmitted pickup ions) and two ring components (ISN pickup ions and NSW pickup ions).
The inverse mirror instability growth rate $\gamma_{mirror}^{-1}$ (normalized by $\Omega_{ci}^{-1}$) is plotted as the orange line in Figure \ref{fig:time}, where we vary the relative number density of the ring components, $n_n / n_i$.
As expected, the inverse growth rate decreases with the ring density.
The instability threshold where $\gamma_{mirror}^{-1} \to \infty$ can be observed from the figure, occuring at about $n_n/n_i \simeq 0.8\%$.
The charge-exchange time scale $\tau$ estimated by Equation \eqref{eq:np} is plotted as the blue line in Figure \ref{fig:time}.
The instability threshold corresponds to a charge-exchange time of $\Omega_{ci}\tau \simeq 3\times 10^4$ or a distance of $L \simeq u_{sw}\tau \simeq 3$ au.
This suggests that there may be a region within a few astronomical units downstream of the termination shock where the newborn pickup ions accumulate while maintaining their initial anisotropy, which may be probed by future in situ observations by New Horizons.
As the newborn pickup ion density continues to increase above the instability threshold deeper in the heliosheath, we expect the anisotropy level to gradually decrease, which is supported by the comparison between our two simulation runs.
The saturated enhancement factor $f_{ani}/f_{iso} \sim 4$ suggested by our 7\% simulation may not be reached until $\sim 20$ au from the termination shock.
Accurate modeling of the observed ENA fluxes will require knowledge of the anisotropic pickup ion distribution along the entire line of sight.
The average effective anisotropy factor may be between the initial value of $\sim 7$ and the saturated value of $\sim 4$ from the simulation.
Finally, we note that the threshold density is much lower in the supersonic solar wind, where the ring velocity is much higher and the ion cyclotron instability is often important.

\section{Summary and conclusions} \label{sec:conclusion}

We investigated the velocity distribution of newborn pickup ions in the heliosheath just downstream of the heliospheric termination shock.
The effects of the neutral solar wind and anisotropy are considered with hybrid kinetic simulations.
The main results are summarized as follows:

1. Even though the density of the neutral solar wind population is much lower than the interstellar neutrals or the hydrogen wall, they may contribute significantly to the observed low-energy ENAs.

2. Mirror instability is expected to be the dominant mode for typical heliosheath condition just downstream of the termination shock.

3. The instability in the heliosheath is much weaker than in the supersonic solar wind. The pickup ions in the heliosheath can maintain a highly anisotropic pitch angle distribution.

4. The combined effects of the neutral solar wind and anisotropy enhance the ion phase-space density at $90^\circ$ pitch angle by a factor of $\sim 120$ based on our simulation.

5. There may be a region within a few astronomical units downstream of the termination shock where the newborn pickup ions do not excite instabilities and the initial ring distribution is preserved due to the low number density. However, the transmitted solar wind turbulence may provide some scattering in this region.

The effects of our results on ENAs can be inferred from the work by \citet{Galli2023}, where the modeled ENA fluxes are separated into contributions from different ion sources, including pickup ions locally created in the heliosheath (denoted as ``IHS PUIs'' in their Figure 3).
We note that the heliosheath pickup ion contribution from the BU (Boston University) model in \citet{Galli2023} has likely been overestimated for a different reason because it used a Maxwellian model, which has a larger thermal spread than a shell-like distribution.
This contributes to the discrepancy between it and the Moscow model, which treats the pickup ions kinetically.
Therefore, the enhancement factor suggested by our simulation is more relevant to the Moscow model, though we note that quantifying the exact enhancement to ENA fluxes needs more careful modeling.
Remarkably, Figure 3 by \citet{Galli2023} shows that the heliosheath-created pickup ion contribution in the Moscow model is a few hundred times lower than the IBEX observations near 100 eV in the upwind direction, and the contribution from transmitted pickup ions (``SSW PUIs'' in the figure) is only a few times higher than heliosheath-created pickup ions.
The enhancement factor of $\sim 120$ suggested by our work, if applicable to the source along an entire line of sight, would close a significant portion of the gap between the model and observations.
The primary source of uncertainty in our estimated enhancement factor is the neutral distribution as it determines the initial newborn pickup ion distribution.
Incorporating the effects of neutral solar wind and anisotropy in ENA models, as well as kinetic neutral modeling, is important for future ENA models.

It is also interesting to note that Voyager observations suggested the presence of compressible mirror-mode structures in the heliosheath \citep{Burlaga2006,Tsurutani2011}.
While anisotropic heating at the termination shock may be a source of these structures \citep{Liu2007}, it was speculated that the heliosheath pickup ions may contribute to the anisotropy \citep{Fichtner2020}.
The mirror instability in our simulation may also help understand in situ observations by Voyagers.

Other sources of turbulence may exist in the heliosheath, such as magnetic reconnection, shear flows, and shocks.
For example, the termination shock can amplify preexisting solar wind turbulence and generate fast magnetosonic turbulence, as supported by models and Voyager observations \citep{Zank2021,Zieger2020}, though shock-generated turbulence is expected to decay away from the shock.
These external turbulence sources are not considered in the present work, and they may cause further isotropization of the pickup ions.
Solar wind turbulence models generally consider pickup ions as the dominant source of turbulence in the outer heliosphere, and this is supported by the observed heating in thermal solar wind ions \citep[e.g.,][]{Zank2018}.
However, other turbulence sources may not be entirely negligible in the distant heliosphere, as indicated by the heating in pickup ions observed by New Horizons \citep{McComas2021}, and the calculation of cosmic ray diffusion \citep{Zhao2017}.
The termination shock and reconnection at the heliospheric current sheet may generate turbulence in the heliosheath, and this may be considered in a future work.

\begin{acknowledgments}
The authors were supported by NASA grant 18-DRIVE18\_2-0029, Our Heliospheric Shield, No.\ 80NSSC22M0164.
MK and MO were partially supported by NASA HGI grant No.\ 80NSSC22K0525.
Suggestions by Lingling Zhao at The University of Alabama in Huntsville are acknowledged.
This research used resources of the National Energy Research Scientific Computing Center, a DOE Office of Science User Facility
supported by the Office of Science of the U.S.\ Department of Energy under Contract No.\ DE-AC02-05CH11231 using NERSC award FES-ERCAP0028261.

A Jupyter Notebook and data for reproducing the figures in this article are available in a Zenodo repository at https://doi.org/10.5281/zenodo.15014461.
\end{acknowledgments}

\bibliography{Du2025}

\begin{thebibliography}{}
\expandafter\ifx\csname natexlab\endcsname\relax\def\natexlab#1{#1}\fi
\providecommand{\url}[1]{\href{#1}{#1}}
\providecommand{\dodoi}[1]{doi:~\href{http://doi.org/#1}{\nolinkurl{#1}}}
\providecommand{\doeprint}[1]{\href{http://ascl.net/#1}{\nolinkurl{http://ascl.net/#1}}}
\providecommand{\doarXiv}[1]{\href{https://arxiv.org/abs/#1}{\nolinkurl{https://arxiv.org/abs/#1}}}

\bibitem[{Baliukin(2024)}]{Baliukin2024}
Baliukin, I. 2024, Fluid Dynamics, 59, 2401, \dodoi{10.1134/S0015462824605175}

\bibitem[{{Baliukin} {et~al.}(2020){Baliukin}, {Izmodenov}, \&
  {Alexashov}}]{Baliukin2020}
{Baliukin}, I.~I., {Izmodenov}, V.~V., \& {Alexashov}, D.~B. 2020, \mnras, 499,
  441, \dodoi{10.1093/mnras/staa2862}

\bibitem[{{Bowers} {et~al.}(2008){Bowers}, {Albright}, {Yin}, {Bergen}, \&
  {Kwan}}]{Bowers2008}
{Bowers}, K.~J., {Albright}, B.~J., {Yin}, L., {Bergen}, B., \& {Kwan},
  T.~J.~T. 2008, Physics of Plasmas, 15, 055703, \dodoi{10.1063/1.2840133}

\bibitem[{{Burlaga} {et~al.}(2006){Burlaga}, {Ness}, \&
  {Acu{\~n}a}}]{Burlaga2006}
{Burlaga}, L.~F., {Ness}, N.~F., \& {Acu{\~n}a}, M.~H. 2006, \apj, 642, 584,
  \dodoi{10.1086/500826}

\bibitem[{{Califano} {et~al.}(2008){Califano}, {Hellinger}, {Kuznetsov},
  {Passot}, {Sulem}, \& {Tr{\'a}Vn{\'\i}{\v{c}}Ek}}]{Califano2008}
{Califano}, F., {Hellinger}, P., {Kuznetsov}, E., {et~al.} 2008, Journal of
  Geophysical Research (Space Physics), 113, A08219,
  \dodoi{10.1029/2007JA012898}

\bibitem[{{Cheng} {et~al.}(2023){Cheng}, {Liu}, {Mousavi}, {Cowee}, {Liu},
  {Wang}, {Wang}, {Zheng}, \& {Zhou}}]{Cheng2023}
{Cheng}, K., {Liu}, K., {Mousavi}, A., {et~al.} 2023, \grl, 50, e2023GL103180,
  \dodoi{10.1029/2023GL103180}

\bibitem[{{Cheng} {et~al.}(2024){Cheng}, {Liu}, {Mousavi}, {Cowee}, {Zheng},
  {Zhou}, {Wang}, {Liu}, \& {Wang}}]{Cheng2024}
---. 2024, \grl, 51, 2024GL111657, \dodoi{10.1029/2024GL111657}

\bibitem[{{Du} {et~al.}(2024){Du}, {Opher}, {Giacalone}, {Guo}, {Richardson},
  \& {Zieger}}]{Du2024}
{Du}, S., {Opher}, M., {Giacalone}, J., {et~al.} 2024, \apj, 974, 210,
  \dodoi{10.3847/1538-4357/ad7374}

\bibitem[{{Fahr} {et~al.}(2016){Fahr}, {Sylla}, {Fichtner}, \&
  {Scherer}}]{Fahr2016}
{Fahr}, H.-J., {Sylla}, A., {Fichtner}, H., \& {Scherer}, K. 2016, Journal of
  Geophysical Research (Space Physics), 121, 8203, \dodoi{10.1002/2016JA022561}

\bibitem[{{Fichtner} {et~al.}(2020){Fichtner}, {Kleimann}, {Yoon}, {Scherer},
  {Oughton}, \& {Engelbrecht}}]{Fichtner2020}
{Fichtner}, H., {Kleimann}, J., {Yoon}, P.~H., {et~al.} 2020, \apj, 901, 76,
  \dodoi{10.3847/1538-4357/abaf52}

\bibitem[{{Florinski} {et~al.}(2016){Florinski}, {Heerikhuisen}, {Niemiec}, \&
  {Ernst}}]{Florinski2016}
{Florinski}, V., {Heerikhuisen}, J., {Niemiec}, J., \& {Ernst}, A. 2016, \apj,
  826, 197, \dodoi{10.3847/0004-637X/826/2/197}

\bibitem[{{Fuselier} {et~al.}(2009){Fuselier}, {Bochsler}, {Chornay}, {Clark},
  {Crew}, {Dunn}, {Ellis}, {Friedmann}, {Funsten}, {Ghielmetti}, {Googins},
  {Granoff}, {Hamilton}, {Hanley}, {Heirtzler}, {Hertzberg}, {Isaac}, {King},
  {Knauss}, {Kucharek}, {Kudirka}, {Livi}, {Lobell}, {Longworth}, {Mashburn},
  {McComas}, {M{\"o}bius}, {Moore}, {Moore}, {Nemanich}, {Nolin}, {O'Neal},
  {Piazza}, {Peterson}, {Pope}, {Rosmarynowski}, {Saul}, {Scherrer}, {Scheer},
  {Schlemm}, {Schwadron}, {Tillier}, {Turco}, {Tyler}, {Vosbury}, {Wieser},
  {Wurz}, \& {Zaffke}}]{Fuselier2009}
{Fuselier}, S.~A., {Bochsler}, P., {Chornay}, D., {et~al.} 2009, \ssr, 146,
  117, \dodoi{10.1007/s11214-009-9495-8}

\bibitem[{{Fuselier} {et~al.}(2021){Fuselier}, {Galli}, {Richardson},
  {Reisenfeld}, {Zirnstein}, {Heerikhuisen}, {Dayeh}, {Schwadron}, {McComas},
  {Elliott}, \& et~al.}]{Fuselier2021}
{Fuselier}, S.~A., {Galli}, A., {Richardson}, J.~D., {et~al.} 2021, \apjl, 915,
  L26, \dodoi{10.3847/2041-8213/ac0d5c}

\bibitem[{{Galli} {et~al.}(2023){Galli}, {Baliukin}, {Kornbleuth}, {Opher},
  {Fuselier}, {Sok{\'o}{\l}}, {Dialynas}, {Dayeh}, {Izmodenov}, \&
  {Richardson}}]{Galli2023}
{Galli}, A., {Baliukin}, I.~I., {Kornbleuth}, M., {et~al.} 2023, \apjl, 954,
  L24, \dodoi{10.3847/2041-8213/aced9b}

\bibitem[{{Gary} \& {Madland}(1988)}]{Gary1988}
{Gary}, S.~P., \& {Madland}, C.~D. 1988, \jgr, 93, 235,
  \dodoi{10.1029/JA093iA01p00235}

\bibitem[{{Giacalone} {et~al.}(2021){Giacalone}, {Nakanotani}, {Zank},
  {K{\`o}ta}, {Opher}, \& {Richardson}}]{Giacalone2021}
{Giacalone}, J., {Nakanotani}, M., {Zank}, G.~P., {et~al.} 2021, \apj, 911, 27,
  \dodoi{10.3847/1538-4357/abe93a}

\bibitem[{{Heerikhuisen} {et~al.}(2016){Heerikhuisen}, {Gamayunov},
  {Zirnstein}, \& {Pogorelov}}]{Heerikhuisen2016}
{Heerikhuisen}, J., {Gamayunov}, K.~V., {Zirnstein}, E.~J., \& {Pogorelov},
  N.~V. 2016, \apj, 831, 137, \dodoi{10.3847/0004-637X/831/2/137}

\bibitem[{{Hellinger} {et~al.}(2017){Hellinger}, {Landi}, {Matteini},
  {Verdini}, \& {Franci}}]{Hellinger2017}
{Hellinger}, P., {Landi}, S., {Matteini}, L., {Verdini}, A., \& {Franci}, L.
  2017, \apj, 838, 158, \dodoi{10.3847/1538-4357/aa67e0}

\bibitem[{{Isenberg} \& {Lee}(1996)}]{Isenberg1996}
{Isenberg}, P.~A., \& {Lee}, M.~A. 1996, \jgr, 101, 11055,
  \dodoi{10.1029/96JA00293}

\bibitem[{{Le} {et~al.}(2021){Le}, {Winske}, {Stanier}, {Daughton}, {Cowee},
  {Wetherton}, \& {Guo}}]{Le2021}
{Le}, A., {Winske}, D., {Stanier}, A., {et~al.} 2021, Journal of Geophysical
  Research (Space Physics), 126, e29125, \dodoi{10.1029/2021JA029125}

\bibitem[{{Lee} \& {Ip}(1987)}]{Lee1987}
{Lee}, M.~A., \& {Ip}, W.~H. 1987, \jgr, 92, 11041,
  \dodoi{10.1029/JA092iA10p11041}

\bibitem[{{Lindsay} \& {Stebbings}(2005)}]{Lindsay2005}
{Lindsay}, B.~G., \& {Stebbings}, R.~F. 2005, Journal of Geophysical Research
  (Space Physics), 110, A12213, \dodoi{10.1029/2005JA011298}

\bibitem[{{Liu} {et~al.}(2012){Liu}, {M{\"o}bius}, {Gary}, \&
  {Winske}}]{Liu2012}
{Liu}, K., {M{\"o}bius}, E., {Gary}, S.~P., \& {Winske}, D. 2012, Journal of
  Geophysical Research (Space Physics), 117, A10102,
  \dodoi{10.1029/2012JA017969}

\bibitem[{{Liu} {et~al.}(2007){Liu}, {Richardson}, {Belcher}, \&
  {Kasper}}]{Liu2007}
{Liu}, Y., {Richardson}, J.~D., {Belcher}, J.~W., \& {Kasper}, J.~C. 2007,
  \apjl, 659, L65, \dodoi{10.1086/516568}

\bibitem[{{McComas} {et~al.}(2009){McComas}, {Allegrini}, {Bochsler},
  {Bzowski}, {Christian}, {Crew}, {DeMajistre}, {Fahr}, {Fichtner}, {Frisch},
  {Funsten}, {Fuselier}, {Gloeckler}, {Gruntman}, {Heerikhuisen}, {Izmodenov},
  {Janzen}, {Knappenberger}, {Krimigis}, {Kucharek}, {Lee}, {Livadiotis},
  {Livi}, {MacDowall}, {Mitchell}, {M{\"o}bius}, {Moore}, {Pogorelov},
  {Reisenfeld}, {Roelof}, {Saul}, {Schwadron}, {Valek}, {Vanderspek}, {Wurz},
  \& {Zank}}]{McComas2009}
{McComas}, D.~J., {Allegrini}, F., {Bochsler}, P., {et~al.} 2009, Science, 326,
  959, \dodoi{10.1126/science.1180906}

\bibitem[{{McComas} {et~al.}(2021){McComas}, {Swaczyna}, {Szalay}, {Zirnstein},
  {Rankin}, {Elliott}, {Singer}, {Spencer}, {Stern}, \& {Weaver}}]{McComas2021}
{McComas}, D.~J., {Swaczyna}, P., {Szalay}, J.~R., {et~al.} 2021, \apjs, 254,
  19, \dodoi{10.3847/1538-4365/abee76}

\bibitem[{{Min} \& {Liu}(2018)}]{Min2018}
{Min}, K., \& {Liu}, K. 2018, \apj, 852, 39, \dodoi{10.3847/1538-4357/aaa0d4}

\bibitem[{{Mousavi} {et~al.}(2020){Mousavi}, {Liu}, \& {Min}}]{Mousavi2020}
{Mousavi}, A., {Liu}, K., \& {Min}, K. 2020, \apj, 901, 167,
  \dodoi{10.3847/1538-4357/abb1a1}

\bibitem[{{Richardson} {et~al.}(2022){Richardson}, {Burlaga}, {Elliott},
  {Kurth}, {Liu}, \& {von Steiger}}]{Richardson2022}
{Richardson}, J.~D., {Burlaga}, L.~F., {Elliott}, H., {et~al.} 2022, \ssr, 218,
  35, \dodoi{10.1007/s11214-022-00899-y}

\bibitem[{{Richardson} {et~al.}(2008){Richardson}, {Kasper}, {Wang}, {Belcher},
  \& {Lazarus}}]{Richardson2008}
{Richardson}, J.~D., {Kasper}, J.~C., {Wang}, C., {Belcher}, J.~W., \&
  {Lazarus}, A.~J. 2008, \nat, 454, 63, \dodoi{10.1038/nature07024}

\bibitem[{{Summerlin} {et~al.}(2014){Summerlin}, {Vi{\~n}as}, {Moore},
  {Christian}, \& {Cooper}}]{Summerlin2014}
{Summerlin}, E.~J., {Vi{\~n}as}, A.~F., {Moore}, T.~E., {Christian}, E.~R., \&
  {Cooper}, J.~F. 2014, \apj, 793, 93, \dodoi{10.1088/0004-637X/793/2/93}

\bibitem[{{Tsurutani} {et~al.}(2011){Tsurutani}, {Echer}, {Verkhoglyadova},
  {Lakhina}, \& {Guarnieri}}]{Tsurutani2011}
{Tsurutani}, B.~T., {Echer}, E., {Verkhoglyadova}, O.~P., {Lakhina}, G.~S., \&
  {Guarnieri}, F.~L. 2011, Journal of Atmospheric and Solar-Terrestrial
  Physics, 73, 1398, \dodoi{10.1016/j.jastp.2010.06.007}

\bibitem[{{Vasyliunas} \& {Siscoe}(1976)}]{Vasyliunas1976}
{Vasyliunas}, V.~M., \& {Siscoe}, G.~L. 1976, \jgr, 81, 1247,
  \dodoi{10.1029/JA081i007p01247}

\bibitem[{{Williams} \& {Zank}(1994)}]{Williams1994}
{Williams}, L.~L., \& {Zank}, G.~P. 1994, \jgr, 99, 19229,
  \dodoi{10.1029/94JA01657}

\bibitem[{{Wu} \& {Davidson}(1972)}]{Wu1972}
{Wu}, C.~S., \& {Davidson}, R.~C. 1972, \jgr, 77, 5399,
  \dodoi{10.1029/JA077i028p05399}

\bibitem[{{Yoon}(1992)}]{Yoon1992}
{Yoon}, P.~H. 1992, Physics of Fluids B, 4, 3627, \dodoi{10.1063/1.860371}

\bibitem[{{Yoon}(2017)}]{Yoon2017}
---. 2017, Reviews of Modern Plasma Physics, 1, 4,
  \dodoi{10.1007/s41614-017-0006-1}

\bibitem[{{Zank} {et~al.}(2018){Zank}, {Adhikari}, {Zhao}, {Mostafavi},
  {Zirnstein}, \& {McComas}}]{Zank2018}
{Zank}, G.~P., {Adhikari}, L., {Zhao}, L.~L., {et~al.} 2018, \apj, 869, 23,
  \dodoi{10.3847/1538-4357/aaebfe}

\bibitem[{{Zank} {et~al.}(2021){Zank}, {Nakanotani}, {Zhao}, {Du}, {Adhikari},
  {Che}, \& {le Roux}}]{Zank2021}
{Zank}, G.~P., {Nakanotani}, M., {Zhao}, L.~L., {et~al.} 2021, \apj, 913, 127,
  \dodoi{10.3847/1538-4357/abf7c8}

\bibitem[{{Zhao} {et~al.}(2017){Zhao}, {Adhikari}, {Zank}, {Hu}, \&
  {Feng}}]{Zhao2017}
{Zhao}, L.~L., {Adhikari}, L., {Zank}, G.~P., {Hu}, Q., \& {Feng}, X.~S. 2017,
  \apj, 849, 88, \dodoi{10.3847/1538-4357/aa932a}

\bibitem[{{Zhao} {et~al.}(2019){Zhao}, {Zank}, \& {Adhikari}}]{Zhao2019}
{Zhao}, L.~L., {Zank}, G.~P., \& {Adhikari}, L. 2019, \apj, 879, 32,
  \dodoi{10.3847/1538-4357/ab2381}

\bibitem[{{Zieger} {et~al.}(2020){Zieger}, {Opher}, {T{\'o}th}, \&
  {Florinski}}]{Zieger2020}
{Zieger}, B., {Opher}, M., {T{\'o}th}, G., \& {Florinski}, V. 2020, Journal of
  Geophysical Research (Space Physics), 125, e28393,
  \dodoi{10.1029/2020JA028393}

\bibitem[{{Zirnstein} {et~al.}(2018{\natexlab{a}}){Zirnstein}, {Kumar},
  {Heerikhuisen}, {McComas}, \& {Galli}}]{Zirnstein2018}
{Zirnstein}, E.~J., {Kumar}, R., {Heerikhuisen}, J., {McComas}, D.~J., \&
  {Galli}, A. 2018{\natexlab{a}}, \apj, 860, 170,
  \dodoi{10.3847/1538-4357/aac3de}

\bibitem[{{Zirnstein} {et~al.}(2018{\natexlab{b}}){Zirnstein}, {Kumar},
  {Heerikhuisen}, {McComas}, \& {Galli}}]{Zirnstein2018b}
---. 2018{\natexlab{b}}, \apj, 865, 150, \dodoi{10.3847/1538-4357/aadb98}

\end{thebibliography}
\bibliographystyle{aasjournal}

\end{document}